\documentclass[12pt,onecolumn]{IEEEtran}

\usepackage{cite}
\usepackage{graphicx}
\usepackage{amsmath}
\usepackage{citesort}
\usepackage{amsfonts}
\usepackage{amssymb,bm,upgreek}

\newcommand{\qed}{\nobreak \ifvmode \relax \else
      \ifdim\lastskip<1.5em \hskip-\lastskip
      \hskip1.5em plus0em minus0.5em \fi \nobreak
      \vrule height0.75em width0.5em depth0.25em\fi}

\begin{document}


\title{Symbol Error Rate of Space-Time Network Coding in Nakagami-$m$ Fading}

\author{Ang Yang, Zesong Fei, Nan Yang, Chengwen Xing, and Jingming Kuang
\thanks{A.~Yang, Z.~Fei, C.~Xing, and J.~Kuang are with the School of Information and Electronics,
Beijing Institute of Technology, Beijing, China (email:taylorkingyang@163.com, feizesong@bit.edu.cn,
chengwenxing@ieee.org, JMKuang@bit.edu.cn).}
\thanks{N.~Yang is with the Wireless and
Networking Technologies Laboratory, CSIRO ICT Centre, Marsfield, NSW
2122, Australia (email: jonas.yang@csiro.au).}}

\maketitle

\begin{abstract}
In this paper, we analyze the symbol error rate (SER) of space-time
network coding (STNC) in a distributed cooperative network over
independent but not necessarily identically distributed (i.n.i.d.)
Nakagami-$m$ fading channels. In this network, multiple sources
communicate with a single destination with the assistance of
multiple decode-and-forward (DF) relays. We first derive new exact
closed-form expressions for the SER with $M$-ary phase shift-keying
modulation ($M$-PSK) and $M$-ary quadrature amplitude modulation
($M$-QAM). We then derive new compact expressions for the asymptotic
SER to offer valuable insights into the network behavior in the high
signal-to-noise ratio (SNR) regime. Importantly, we demonstrate that
STNC guarantees full diversity order, which is determined by the
Nakagami-$m$ fading parameters of all the channels but independent
of the number of sources. Based on the new expressions, we examine
the impact of the number of relays, relay location, Nakagami-$m$
fading parameters, power allocation, and nonorthogonal codes on the
SER.
\end{abstract}



\IEEEpeerreviewmaketitle

\newpage

\section{Introduction}

Cooperative communication has been recognized as a promising
low-cost solution to combat fading and to extend coverage in
wireless networks
\cite{Naka_df:A_Sendonaris_cooper,Naka_df:David_Tse_cooper}. The key
idea behind this solution is to employ relays to receive and
transmit the source information to the destination, which generates
a virtual multiple-input and multiple-output (MIMO) system to
provide spatial cooperative diversity
\cite{rvp:Phee_Yeoh,rvq:Yuan_AF_beamforming}. Apart from diversity,
throughput enhancement is another critical challenge for wireless
networks. Against this background, network coding (NC) is proposed
as a potentially powerful tool to enable efficient information
transmission, where data flows coming from multiple sources are
combined to increase throughput
\cite{MUNC:NC_flow,MUNC:NC_flow_2,rvq:Jun_Li}.

Recently, the joint exploitation of cooperative diversity and NC has
become a primary design concern in distributed networks with
multiple users and multiple relays
\cite{MUNC:Ming_Xiao,Naka_df:Ming_Xiao_multi-user_LNC,MUNC:Yuan,Naka_df:Ming_Xiao_NC_backhaul}.
Motivated by this, \cite{MUNC:Ming_Xiao} investigated various sink
network decoding approaches for the network with intermediate node
encoding. In \cite{Naka_df:Ming_Xiao_multi-user_LNC}, linear network
coding (LNC) was applied in distributed uplink networks to
facilitate the transmission of independent information from multiple
users to a common base station. In \cite{MUNC:Yuan}, low-density
parity-check (LDPC) code and NC was jointly designed for a
multi-source single-relay FDMA system over uniform phase-fading
Gaussian channels. In \cite{Naka_df:Ming_Xiao_NC_backhaul},
cooperative network coding strategies were proposed for a
relay-aided two-source two-destination wireless network with a
backhaul connection between the sources.




In order to increase the capacity and transmission reliability of
wireless cooperative networks, multiple antennas are deployed to
gain the merits of MIMO processing techniques
\cite{Naka_df:ST_Yuan,Naka_df:Eric_ST,Naka_df:Xiqi_Gao_Turbo_rec_STBC,Naka_df:ST_Yuan_2,Naka_MIMOR:Louie,Naka_MIMOR:Yeoh}.
In this strategy, distributed space-time coding (DSTC) was proposed
to further boost network performance, where the antennas at the
distributed relays are utilized as transmit antennas to generate a
space-time code for the receiver
\cite{Naka_df:DSTC_laneman,Jing:Distributed_STC_in_WSN}. A
differential DSTC was proposed in \cite{Jing:DST_differential} to
eliminate the requirement of channel information at the relays and
the receiver. The combined benefits of maximum-ratio combining (MRC)
and DSTC were investigated in
\cite{Jing:Combination_of_MRC_and_Distributed_Space-Time}. In
\cite{Naka_df:ShiJin_PA_DSTC_Two-way}, the impact of DSTC in two-way
amplify-and-forward relay channels was characterized.


One of the principal challenges in distributed cooperative networks
is to leverage the benefits from both NC and DSTC. A promising
solution that addresses this challenge is space-time network coding
(STNC), which was proposed in \cite{MUNC:Ray_Liu_STNC}.
Fundamentally, STNC combines the information from different sources
at a relay, which involves the concept of NC. Moreover, STNC
transmits the combined signals in several time slots using a set of
relays, which involves the concept of DSTC. Based on these, STNC
achieves spatial diversity with low transmission delay under the
impact of imperfect frequency and timing synchronization. We note
that in \cite{MUNC:Ray_Liu_STNC}, the performance of STNC was
evaluated for Rayleigh fading channels, where no closed-form
expression was presented.


In this paper, we consider a distributed cooperative network using
STNC over independent but not necessarily identically distributed
(i.n.i.d.) Nakagami-$m$ fading channels, which generalizes the
result in \cite{MUNC:Ray_Liu_STNC}. In this network, multiple
sources communicate with a single destination with the assistance of
multiple relays. Here, we focus on decode-and-forward (DF) protocol
at the relays, which arises from the fact that this protocol has
been successfully deployed in practical wireless standards, e.g.,
3GPP Long Term Evolution (LTE) and IEEE 802.16j WiMAX \cite{DF_LTE}.
Different from \cite{MUNC:Ray_Liu_STNC}, we examine two fundamental
questions as follows: ``\emph{1) What is the impact of STNC on the
symbol error rate (SER) in general Nakagami-$m$ fading channels?}''
and ``\emph{2) Can we provide closed-form expressions for the SER in
Nakagami-$m$ fading to alleviate the burden of Monte Carlo
simulations?}'' The rationale behind these questions is that
Nakagami-$m$ fading covers a wide range of typical fading scenarios
in realistic wireless applications via the $m$ parameter. Notably,
Nakagami-$m$ fading encompasses Rayleigh fading ($m = 1$) as a
special case \cite{MUNC:nakagami}. To tackle these questions, we
first derive new closed-form expressions for the exact SER, which
are valid for multiple phase shift-keying modulation ($M$-PSK) and
$M$-ary quadrature amplitude modulation ($M$-QAM). To further
provide valuable insights at high signal-to-noise ratios (SNRs), we
derive new compact expressions for the asymptotic SER, from which
the diversity gain is obtained. Specifically, it is demonstrated
that the diversity order is determined by the Nakagami-$m$ fading
parameters of all the channels, but independent of the number of
sources. Various numerical results are utilized to examine the
impact of the number of relays, relay location, Nakagami-$m$ fading
parameters, power allocation, and nonorthogonal codes on the SER.
Importantly, it is shown that nonorthogonal codes provide higher
throughput than orthogonal codes, while guaranteeing full diversity
over Nakagami-$m$ fading channels. Our analytical expressions are
substantiated via Monte Carlo simulations.

\section{System Model}

\begin{figure}[!t]
\centering
\includegraphics[width=4in]{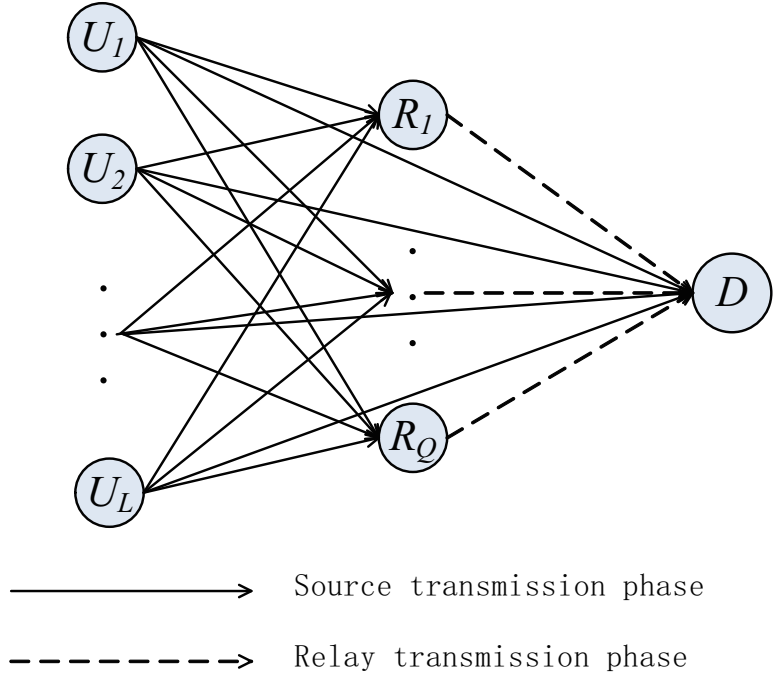}
\caption{System model.} \label{system_stnc.eps}
\end{figure}

Fig.~\ref{system_stnc.eps} depicts a distributed cooperative network
where $L$ sources, $U_1, U_2, \ldots, U_L$, transmit their own
information to a common destination $D$ with the aid of $Q$ relays,
$R_1, R_2, \ldots, R_Q$. In this network, each node is equipped with
a single antenna. We consider a practical and versatile operating
scenario where the source-relay, the relay-destination, and the
source-destination channels experience i.n.i.d. Nakagami-$m$ fading.
As such, we denote the Nakagami-$m$ fading parameter between $U_l$
and $R_q$ as $m_{lq}$, the Nakagami-$m$ fading parameter between
$R_q$ and $D$ as $m_{qd}$, and the Nakagami-$m$ fading parameter
between $U_l$ and $D$ as $m_{ld}$. We further denote the channel
coefficient between $U_{l}$ and $R_{q}$ as $h_{lq}$, the channel
coefficient between $R_{q}$ and $D$ as $h_{qd}$, and the channel
coefficient between $U_{l}$ and $D$ as $h_{ld}$, where
$1\leq{}l\leq{}L$ and $1\leq{}q\leq{}Q$. Throughout this paper, we
define the variances of these channel coefficients as
$h_{\phi\varphi}\sim\mathcal{CN}\left(0,d_{\phi\varphi}^{-\alpha}\right)$,
where $\phi\in\left\{l,q\right\}$, $\varphi\in\left\{q,d\right\}$,
and $\phi\neq\varphi$. Here, we incorporate the path loss in the
signal propagation such that $d_{\phi\varphi}$ denotes the distance
between $\phi$ and $\varphi$ and $\alpha$ denotes the path loss
exponent.



\begin{figure}[!t]
\centering
\includegraphics[width=5.5in]{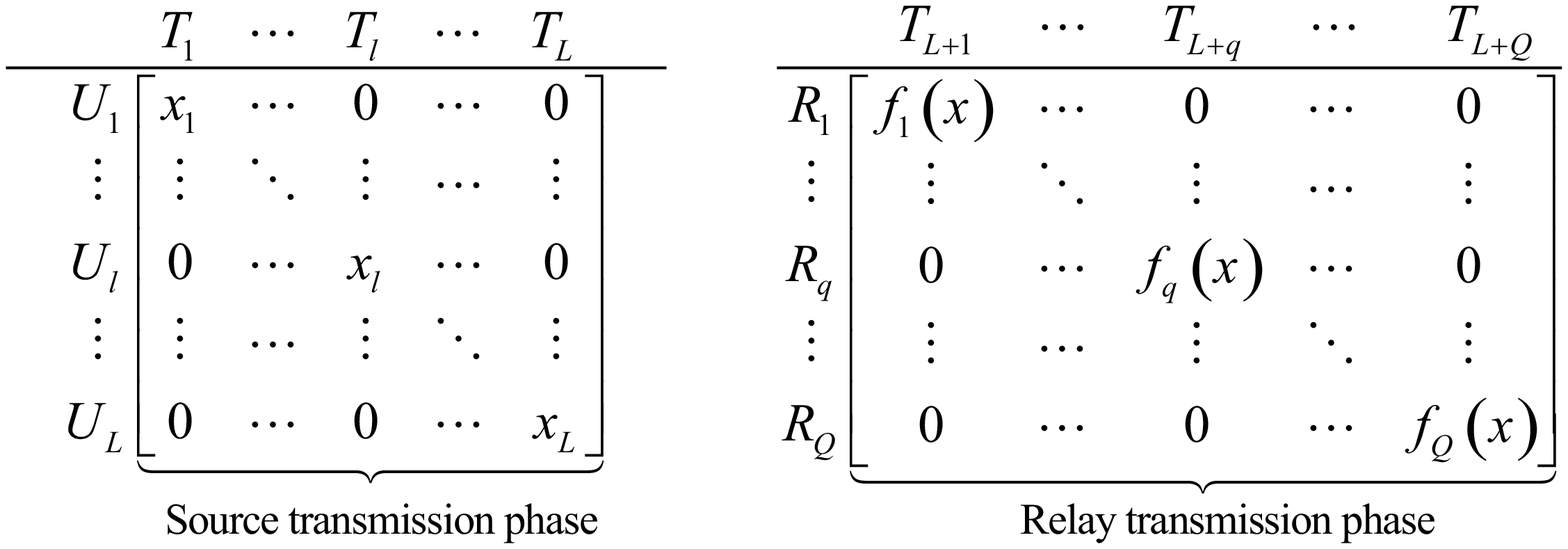}
\caption{A framework of space-time network coding.} \label{stnc.eps}
\end{figure}

In this network, the STNC transmission between the sources and the
destination is divided into two consecutive phases: 1) source
transmission phase and 2) relay transmission phase. In the source
transmission phase, the sources transmit their symbols in the
designated time slots. In this phase, the relays receive a set of
overheard symbols from the sources. In the relay transmission phase,
each relay encodes the set of overheard symbols to a single signal
and then transmits it to the destination in its designated time
slot. As illustrated in Fig.~\ref{stnc.eps}, $(L+Q)$ time slots are
required to complete the STNC transmission to eliminate the
detrimental effects of imperfect synchronization on any
point-to-point transmission in this network at any time slot.


We proceed to detail the transmission in the two phases, as follows:

In the source transmission phase, the signals received at the
destination from $U_l$ in the time slot $l$ is given by
\begin{equation}\label{STNC_source transmit 1}
y_{ld}\left(t\right)=h_{ld}\sqrt{P_{l}}x_{l}s_{l}\left(t\right)+w_{ld}\left(t\right),
\end{equation}
where $P_{l}$ denotes the transmit power at $U_l$, $x_l$ denotes the
symbol transmitted by $U_l$, $s_l\left(t\right)$ denotes the
spreading code of $x_l$, and $w_{ld}$ is the additive white Gaussian
noise (AWGN) with zero mean and the variance of $N_0$. The cross
correlation between $s_p\left(t\right)$ and $s_q\left(t\right)$ are
expressed as
$\rho_{pq}=\left\langle{s_p\left(t\right),s_q\left(t\right)}\right\rangle$,
where $\left\langle{f\left(t\right),g\left(t\right)}\right\rangle
\triangleq\frac{1}{T}\int_0^T{f\left(t\right)g^{\ast}\left(t\right)dt}$
is the inner product between $f\left(t\right)$ and $g\left(t\right)$
during the symbol interval $T$. Moreover, we assume that
$\rho_{ll}=\left\|{s_l\left(t\right)}\right\|^{2}=1$. The signals
received at $R_q$ from $U_l$ is given by
\begin{equation} \label{STNC_source transmit 2}
y_{lq}\left(t\right)=h_{lq}\sqrt{P_{l}}x_{l}s_{l}\left(t\right)+w_{lq}\left(t\right),
\end{equation}
where $w_{lq}\left(t\right)$ is AWGN with zero mean and the variance
of $N_0$.

In the relay transmission phase, the signal received at the
destination from $R_q$ is given by
\begin{equation}\label{STNC_relay transmit 1}
y_{qd}\left(t\right)={h_{qd}}\underbrace{\sum_{l=1}^{L}{{\beta
_{ql}}\sqrt {{P_{ql}}} {x_l}{s_l}\left(t\right)} }_{{f_q}\left( x
\right)} + {w_{qd}}\left(t\right),
\end{equation}
where $P_{ql}$ denotes the transmit power at $R_q$ and $w_{qd}$ is
AWGN with zero mean and the variance of $N_0$. In (\ref{STNC_relay
transmit 1}), the scalar $\beta_{ql}$ denotes the state whether
$R_q$ decodes $x_l$ correctly. Specifically, $\beta_{ql}$ is equal
to $1$ if $R_{q}$ decodes $x_{l}$ correctly, but $0$ otherwise.


For the detection of the received signals at the destination, we
assume that the full knowledge of the channel state information are
available at the receivers with the aid of a preamble in the
transmitted signal. We also assume that the destination has the
detection states at the relays, which can be obtained via an
indicator in the relaying signal. At the destination, the spreading
codes $s_{l}$ is employed such that the information symbols $x_{l}$
is separated from $y_{ld}$ and $y_{qd}$, where $l\in\{1,\cdots,L\}$.
For any desired symbol $x_l$, the destination combines the
information of $x_l$ from $U_l$ and the $Q$ relays using maximum
ratio combining (MRC). Therefore, the instantaneous signal-to-noise
ratio (SNR) of $x_l$ is expressed as \cite{MUNC:Ray_Liu_STNC}
\begin{equation}\label{STNC_SNR 1}
\gamma_l  = \frac{{P_l \left| {h_{ld} } \right|^2 }}{{N_0 }} +
\sum_{q = 1}^Q {\frac{{\beta_{ql} P_{lq} \left| {h_{qd} } \right|^2
}}{{N_0 \varepsilon _l }}},
\end{equation}
where $\varepsilon _l$ is the $l$th diagonal element of matrix
${\bf{R}}^{-1}$ associated with symbol $x_l$ and $\bf{R}$ is given
by
\begin{equation}\label{STNC_SNR 2}
{\bf{R}} = \left[{\begin{array}{*{20}c}
1 & {\rho _{21} } &  \cdots  & {\rho _{Q1} }  \\
{\rho _{12} } & 1 &  \cdots  & {\rho _{Q2} }  \\
\vdots  &  \vdots  &  \ddots  &  \vdots   \\
{\rho _{1Q} } & {\rho _{2Q} } &  \cdots  & 1  \\
\end{array}}
\right].
\end{equation}



To facilitate the performance analysis in the following section, we
re-express (\ref{STNC_SNR 1}) as a unitary expression given by
\begin{equation}\label{STNC_SNR}
\gamma_l=c_0\left|{h_0}\right|^{2}+\sum_{q=1}^{Q}\beta_{ql}c_q\left|{h_q}\right|^{2},
\end{equation}
where $c_0=P_{l}d_{dl}^{-\alpha}/N_{0}$ denotes the equivalent SNR
at $D$ received from $U_l$, $h_0$ denotes the unitary Nakagami-$m$
fading coefficient between $U_l$ and $D$ with variance one,
$c_q=P_{l}d_{ql}^{-\alpha}/N_{0}\varepsilon_{l}$ denotes the $q$th
equivalent SNR received at $D$ from $R_q$, and $h_q$ denotes the
unitary Nakagami-$m$ fading coefficient between $R_q$ and $D$ with
variance one.

\section{SER Analysis over Nakagami-$m$ Fading Channel}

In this section, we first derive new closed-form expressions for the
exact SER with $M$-PSK and $M$-QAM. We then derive new compact
expressions for the asymptotic SER, which will allow us to examine
the network behavior in the high SNR regime.

\subsection{Exact SER}

In DF protocol, $\beta_{ql}$ denotes the decoding state at $R_q$
associated with $x_l$. Based on the values of all $\beta_{ql}$'s, we
define a decimal number as
$S_{l}=\left[\beta_{1l}~\beta_{2l}~\cdots~\beta_{Ql}\right]_{2}$ to
represent one of $2^Q$ network decoding states at $Q$ relays
associated with $x_l$. Since all the channels in this network are
mutually independent, the events that whether $R_q$ correctly
decodes the received signal are independent. It follows that
$\beta_{ql}$'s are independent Bernoulli random variables, the
distribution of which is written as
\begin{equation}\label{STNC_G(beta)}
G\left({\beta_{ql}}\right)=\left\{{\begin{array}{ll}
{1-{\rm{SER}}_{ql} ,} & {{\rm{if}}~{\beta_{ql}=1}}  \\
{{\rm{SER}}_{ql},} & {{\rm{otherwise,}}}  \\
\end{array}}
\right.
\end{equation}
where ${\rm{SER}}_{ql}$ denotes the SER of detecting $x_l$ at $R_q$.
Therefore, the joint probability of a particular combination of
$x_l$ in $S_{l}$ is written as
\begin{equation}\label{STNC_Pr Sq}
{\Pr}\left(S_{l}\right)=\prod_{q=1}^{Q}{G\left(\beta_{ql}\right)}.
\end{equation}
Applying Bayesian rule, the SER of detecting $x_l$ at $D$ is derived
as
\begin{align}\label{SER naka final}
{\rm{SER}}_{l}=\sum_{S_{l}=0}^{2^{Q}-1}{{\rm{SER}}_{\gamma_{l|S_l}}\Pr\left({S_l}\right)}
=\sum_{S_{l}=0}^{2^{Q}-1}{{\rm{SER}}_{\gamma_{l|S_l}}\prod_{q=1}^Q{G\left({\beta_{ql}}\right)}},
\end{align}
where ${{\rm{SER}}_{\gamma _{l|S_l } }}$ denotes the SER of
detecting $x_l$ at $D$ conditioned on $S_l$. To facilitate the
calculation of (\ref{SER naka final}), we present the exact
closed-form results for ${{\rm{SER}}_{\gamma_{l|S_l}}}$ and
${G\left({\beta_{ql}}\right)}$, as follows.

\subsubsection{Exact Results for ${\rm{SER}_{\gamma_{l|S_l}}}$}

We commence the derivation of ${\rm{SER}_{\gamma_{l|S_l}}}$ by
presenting the PDF of $\gamma_{l|S_{l}}$,
$f_{\gamma_{l|S_{l}}}\left(v\right)$. If there are $N$ ``$1$''
elements in one set $S_{l}$, let $a_1, a_2, \cdots, a_N \in
\left\{c_1, c_2, \cdots, c_Q \right\}$ denote the equivalent SNRs of
the $N$ relays which decode $x_l$ successfully, where $N \le Q$. We
then define $a_0 =c_0$ to make the source equivalent to the zeroth
relay. As such, the SNR at $D$ to one desired symbol $x_l$ can be
rewritten as
\begin{equation}\label{STNC_SNR S_l}
\gamma _{l|S_l }=a_0 \left| {h_0 } \right|^2 + \sum_{n = 1}^N {a_n
\left| {h_n } \right|^2 } = \sum_{n = 0}^N {\underbrace {a_n \left|
{h_n } \right|^2 }_{Y_n}}.
\end{equation}
According to \cite{MUNC:nakagami}, the PDF of $Y_n$ in Nakagami-$m$
fading is given by
\begin{equation}
\label{Y_n Naka} f_{Y_n } \left( y \right) = \frac{{{m_n} ^{m_n}
y^{m_n - 1} }}{{{a_n} ^{m_n} \Gamma \left( m_n \right)}}\exp \left(
{ - \frac{{m_n y}}{{a_n }}} \right),
\end{equation}
where $m_n$ is the Nakagami-$m$ fading parameter between the $n$th
successful relay and the destination. For example, if the first and
the second relays out of three relays successfully decode the
information from $U_l$, we have $m_1=m_{1d}$ and $m_2=_{2d}$. Using
Fourier transform together with \cite[eq.
(3.351.3)]{rvq:Table_of_Integrals}, the characteristic function (CF)
of $Y_n$ is calculated as
\begin{equation}\label{Y_n  CF Naka}
\begin{split}
C_{Y_n } \left( u \right) &= \int_{-\infty
}^\infty{f_{Y_n}\left(y\right)e^{juy}dy}
=\left( {1 - \frac{{jua_n }}{m_n}} \right)^{ - m_n}.
\end{split}
\end{equation}
Given that $\gamma_{l|S_l}$ is the sum of $Y_n$'s, the CF of $\gamma
_{l|S_l }$ is obtained as
\begin{equation}\label{CF SNR naka}
\begin{split}
C_{\gamma _{l|S_l } }\left( u \right) &= \prod\limits_{n = 0}^N
{C_{Y_n } \left( u \right)}=\prod\limits_{n = 0}^N {\left( {1 -
\frac{{jua_n }}{m_n}} \right)^{ - m_n}}.
\end{split}
\end{equation}
Applying inverse Fourier transform, the PDF of $\gamma _{l|S_l } $
is derived as
\begin{equation}
\label{PDF SNR 1 naka}
\begin{split}
f_{\gamma _{l|S_l } }  \left( v \right) &= \frac{1}{{2\pi }}\int_{ - \infty }^\infty  {C_{\gamma _{l|S_l }}   \left( u \right)e^{ - juv} du}  \\
&= \frac{1}{{2\pi }}\int_{ - \infty }^\infty  {\underbrace {\left(
{\prod\limits_{n = 0}^N {\left( {1 - \frac{{jua_n }}{m_n}} \right)^{
- m_n} } } \right)e^{ - juv} }_{g\left( u \right)}du}.
\end{split}
\end{equation}

We next seek the solution for $g\left(u\right)$. Due to the
randomicity of the wireless channels, we note that $g(u)$ has $N+1$
different poles $z_0=-jm_0/a_0 $, $z_1=-jm_1/a_1$, $\cdots$,
$z_N=-jm_N/a_N$ in complex field. As such, based on residue theorem
\cite{rvq:Complex_analysis}, the residue of $k^{\rm{th}}$ pole of
$g(u)$ in complex field can be expressed as
\begin{align}\label{residue of kth pole naka}
{\rm{Res}}\left[{g\left({z_k}\right),z_k}\right]
=&\frac{1}{{\left( {m_k - 1} \right)!}}\mathop {\lim }\limits_{u \to z_k } \frac{{d^{m_k - 1} }}{{du^{m_k - 1} }}\left[ {\left( {u - z_k } \right)^{m_k} g\left( u \right)} \right]\notag\\
=&\frac{{{m_k}^{m_k} }}{{(-j)^{m_k} {a_k}^{m_k} \left( {m_k - 1}
\right)!}}\mathop {\lim }\limits_{u \to z_k } \frac{{d^{m_k - 1}
}}{{du^{m_k - 1} }}\left[ {e^{ - juv} \left( {\prod\limits_{n = 0,n
\ne k}^N {\left( {1 - \frac{{jua_n }}{m_n}} \right)^{ - m_n} } }
\right)}\right].
\end{align}
As per the general Leibniz's rule, we derive
${\rm{Res}}\left[{g\left({z_k}\right),z_k}\right]$ in (\ref{residue
of kth pole naka}) as
\begin{align}\label{residue of kth pole naka 2 1}
&{\rm{Res}}\left[{g\left({z_k}\right),z_k}\right]\notag\\
=&\frac{{{m_k}^{m_k} }}{{(-j)^{m_k} a_k^{m_k}\left( {m_k - 1} \right)!}}\mathop {\lim }\limits_{u \to z_k } \sum_{i = 0}^{m_k - 1} {\left[ {\left( { - jv} \right)^{m_k - 1 - i} e^{ - juv} } \right]}
\sum_{i_0  = 0}^i {\sum_{i_1  = 0}^{i_0 } { \cdots \sum_{i_{k - 1}  = 0}^{i_{k - 2} } {\sum_{i_{k + 1}  = 0}^{i_{k - 1} } { \cdots \sum_{i_{N - 1}  = 0}^{i_{N - 2} } } } } }\notag\\
&\times{{m_k - 1}\choose i}{i\choose i_{0}}{i_{0}\choose
i_{1}}\cdots{i_{k-2}\choose i_{k-1}}{i_{k-1}\choose
i_{k+1}}\cdots{i_{N-2}\choose i_{N-1}}\left[ {\frac{{d^{i_{N - 1} } }}{{du^{i_{N - 1} } }}\left( {1 - \frac{{jua_N }}{m_N}} \right)^{ - m_N} } \right]\notag\\
&\times \left[ {\frac{{d^{i_{N - 2}  - i_{N - 1} } }}{{du^{i_{N - 2}
- i_{N - 1} } }} \left({1 - \frac{{jua_{N - 1} }}{m_{N-1}}}
\right)^{ - m_{N-1}} } \right] \cdots \left[ {\frac{{d^{i_{k - 1}  -
i_{k + 1} } }}{{du^{i_{k - 1}  - i_{k + 1} } }}
\left({1 - \frac{{jua_{k + 1} }}{m_{k+1}}} \right)^{ - m_{k+1}} } \right]\notag\\
&\times\left[ {\frac{{d^{i_{k - 2}  - i_{k - 1} } }}{{du^{i_{k - 2}
- i_{k - 1} } }}\left( {1 - \frac{{jua_{k - 1} }}{m_{k-1}}}
\right)^{ - m_{k-1}} } \right] \cdots \left[ {\frac{{d^{i_0  - i_1 }
}}{{du^{i_0  - i_1 } }}\left( {1 - \frac{{jua_1 }}{m_1}} \right)^{ -
m_1} } \right]\left[ {\frac{{d^{i - i_0 } }}{{du^{i - i_0 } }}\left(
{1 - \frac{{jua_0 }}{m_0}} \right)^{ - m_0} }\right].
\end{align}
Upon close observation, we simplify (\ref{residue of kth pole naka 2
1}) as
\begin{equation}\label{residue of kth pole naka 2}
{\rm{Res}}\left[{g\left({z_k}\right),z_k}\right]=\sum_{i = 0}^{m_k -
1} {jB_{N,k,i} } v^{m_k - 1 - i} \exp \left( { - \frac{{m_k v}}{{a_k
}}}\right),
\end{equation}
where $B_{N,k,i}$ is defined as
\begin{align}\label{B_k naka}
&B_{N,k,i}\\=&\frac{{{m_k}^{m_k} }{(-1)}^{-i}}{{a_k^{m_k} \left( {{m_k} - 1} \right)!}}\sum_{i_0  =
0}^i {\sum_{i_1  = 0}^{i_0 } {\cdots \sum_{i_{k - 1}  = 0}^{i_{k - 2} } {\sum_{i_{k + 1}  = 0}^{i_{k -
1} } { \cdots \sum_{i_{N - 1}  = 0}^{i_{N - 2} }} } } }{{m_k - 1}\choose i}{i\choose
i_{0}}{i_{0}\choose i_{1}}\cdots{i_{k-2}\choose i_{k-1}}{i_{k-1}\choose
i_{k+1}}\cdots{i_{N-2}\choose i_{N-1}}\notag\\
&\times\left({\frac{{a_{N} }}{{m_N }} }\right)^{i_{N - 1}}\left({\frac{{a_{N-1} }}{{m_{N-1} }} }\right) ^{i_{N - 2}  - i_{N - 1} }  \cdots \left({\frac{{a_{k+1} }}{{m_{k+1} }} }\right) ^{i_{k - 1}  - i_{k + 1} } \left({\frac{{a_{k-1} }}{{m_{k-1} }} }\right) ^{i_{k - 2}  - i_{k + 1} }  \cdots \left({\frac{{a_{1} }}{{m_{1} }} }\right) ^{i_0  - i_1 } \left({\frac{{a_{0} }}{{m_{0} }} }\right) ^{i - i_0 }\notag\\
&\times \left( m_N \right)_{i_{N - 1} } \left( m_{N-1} \right)_{i_{N - 2}  - i_{N - 1} }  \cdots \left( m_{k+1} \right)_{i_{k - 1}  - i_{k + 1} } \left( m_{k-1} \right)_{i_{k - 2}  - i_{k - 1} }  \cdots \left( m_1 \right)_{i_0  - i_1 } \left( m_0 \right)_{i - i_0 }\notag\\
&\times \left( {1 - \frac{{m_k a_N }}{{m_N a_k }}} \right)^{ - m_{N} - i_{N - 1} } \left( {1 - \frac{{m_k a_{N - 1} }}{{m_{N-1} a_k }}} \right)^{ - m_{N-1} - i_{N - 2}  + i_{N - 1} }  \cdots \left( {1 - \frac{{m_k a_{k + 1} }}{{m_{k+1} a_k }}} \right)^{ - m_{k+1} - i_{k - 1}  + i_{k + 1} }\notag\\
&\times \left( {1 - \frac{{m_k a_{k - 1} }}{{m_{k-1} a_k }}} \right)^{ - m_{k-1} - i_{k - 2}  + i_{k -
1} }  \cdots \left( {1 - \frac{{m_k a_1 }}{{m_1 a_k }}} \right)^{ - m_1 - i_0  + i_1 } \left( {1 -
\frac{{m_k a_0 }}{{m_0 a_k }}} \right)^{ - m_0 - i + i_0},
\end{align}
and $\left( m_n \right)_i = \Gamma \left( {m_n + i} \right)\slash
\Gamma \left( m_n \right)$ is the Pochmann symbol. Based on the
residues of the poles, we confirm that $g(u)$ is available for the
residue theorem. Specifically, using the residue of $k^{\rm{th}}$
pole of $g(u)$ in complex field, we obtain
\begin{equation}\label{residue of all pole naka}
\int_{ - \infty }^\infty  {g\left( u \right)} du =  - 2\pi j\sum_{k
= 0}^N {{\mathop{\rm Res}\nolimits} \left[ {g\left( {z_k }
\right),z_k } \right]}.
\end{equation}
The proof of (\ref{residue of all pole naka}) is shown in
Appendix~\ref{app_Residue}. Based on (\ref{residue of kth pole naka
2}) and (\ref{residue of all pole naka}), the PDF of $\gamma _{l|S_l
} $ in (\ref{PDF SNR 1 naka}) is derived as
\begin{equation}\label{pdf SNR 2 naka}
f_{\gamma _{l|S_l } } \left( v \right) = \sum_{k = 0}^N {\sum_{i =
0}^{m_k - 1} {B_{N,k,i} } v^{m_k - 1 - i} \exp \left( { - \frac{{m_k
v}}{{a_k }}} \right)}.
\end{equation}

With the aid of $f_{\gamma_{l|S_{l}}}\left(v\right)$ in (\ref{pdf
SNR 2 naka}), we are capable to derive ${\rm{SER}}_{\gamma _{l|S_{l}
} }$ for $M$-PSK and $M$-QAM. First, ${\rm{SER}}_{\gamma
_{l|S_{l}}}$ for $M$-PSK is derived as
\begin{align}\label{SER naka}
{\rm{SER}}_{\gamma_{l|S_{l}},{\textrm{MPSK}}}&= \frac{1}{\pi }\int_0^{\left( {M - 1} \right)\pi /M} {\int_0^\infty  {\alpha\exp \left( { - \frac{{b\gamma }}{{\sin ^2 \theta }}} \right)f_{\gamma_{l|S_{l} } } \left( \gamma  \right)} d\gamma d\theta }\notag \\
&=\frac{\alpha}{\pi }\sum_{k = 0}^N {\sum_{i = 0}^{m_k - 1}
{B_{N,k,i} } } \int_0^{\left( {M - 1} \right)\pi /M} {\int_0^\infty
{\gamma ^{m_k - 1 - i} \exp \left( {\left( { - \frac{b}{{\sin ^2
\theta }}-\frac{m_k}{{a_k }}} \right)\gamma } \right)} d\gamma
d\theta },
\end{align}
where $\alpha$ and $b$ are modulation specific constants. For
$M$-PSK, $\alpha=1$ and $b=\sin^2(\pi/M)$. With the aid of \cite[eq.
(3.351.3)]{rvq:Table_of_Integrals} and \cite[eq.
(5A.17)]{rvq:digital_com_over_fading_channels}, (\ref{SER naka}) is
derived as
\begin{align}\label{SER naka_PSK}
&{\rm{SER}}_{\gamma_{l|S_{l}},{\textrm{MPSK}}}\notag\\
=&\frac{\alpha}{\pi}\sum_{k = 0}^N {\sum_{i = 0}^{m_k - 1}
{B_{N,k,i} } } \Gamma \left( {m_k - i} \right)\int_0^{\left( {M - 1}
\right)\pi /M}{\left(
{\frac{b}{{\sin ^2 \theta }} + \frac{m_k}{{a_k }}} \right)^{ - m_k + i} d\theta}\notag\\
=&\alpha\sum_{k = 0}^N {\sum_{i = 0}^{m_k - 1} { B_{N,k,i} } }
\Gamma \left( {m_k - i} \right)\left( {\frac{{a_k }}{m_k}}
\right)^{m_k - i} \left[ \frac{{M - 1}}{M} - \frac{1}{\pi }\sqrt
{\frac{{a_k b}}{{a_k b + m_k}}} \left( \left( {\frac{\pi }{2} + \tan
^{-1}\omega } \right)\sum_{p = 0}^{m_k - i - 1}{2p \choose p}\right.\right.\notag\\
&\left.\left.\times\left(4\left( {1 + \frac{{a_k
b}}{m_k}}\right)\right)^{-p}+\sin\left({\tan ^{ - 1} \omega }
\right)\sum_{p = 1}^{m_k - i - 1}\sum_{t = 1}^{p}T_{p,t}
\left(1+\frac{a_k b}{m_k}\right)^{-p}\left({\cos \left( {\tan ^{ -
1} \omega } \right)}\right)^{2\left( {p - t} \right) +
1}\right)\right],
\end{align}
where $\omega  = \sqrt {\frac{{a_k b}}{{a_k b + m_k}}} \cot
\frac{\pi }{M}$ and $T_{p,t} = \frac{{2p \choose p}}{{2(p-t) \choose
{p-t}}4^t\left[{2\left(p - t\right) + 1} \right]}$.

We then derive ${\rm{SER}}_{\gamma _{l|S_{l}}}$ for $M$-QAM as
\begin{align}\label{SER naka_2}
{\rm{SER}}_{\gamma_{l|S_{l}},{\textrm{MQAM}}}=&\frac{4}{\pi
}\int_0^{\pi /2} {\int_0^\infty  {\alpha \exp \left( { -
\frac{{b\gamma }}{{{{\sin }^2}\theta }}} \right){f_{{\gamma
_{l|{S_{l}}}}}}\left( \gamma  \right)} d\gamma d\theta }\notag\\
&-\frac{4}{\pi }\int_0^{\pi /4} {\int_0^\infty  {\alpha^2 \exp
\left( { - \frac{{b\gamma }}{{{{\sin }^2}\theta }}}
\right){f_{{\gamma _{l|{S_{l}}}}}}\left( \gamma  \right)} d\gamma
d\theta },
\end{align}
where $\alpha=4\left( {1 - 1/\sqrt M } \right)$ and
$b=3/\left(2\left( {M - 1} \right)\right)$. Calculating the
integrals in (\ref{SER naka_2}),
${\rm{SER}}_{\gamma_{l|S_{l}},{\textrm{MQAM}}}$ is derived as
\begin{align}\label{SER naka_QAM}
&{\rm{SER}}_{\gamma_{l|S_{l}},{\textrm{MQAM}}}\notag\\
=&4\alpha\sum\limits_{k = 0}^N {\sum\limits_{i = 0}^{{m_{_k}} - 1}
{{B_{k,i}}} } \Gamma \left( {{m_k} - i} \right){\left(
{\frac{{{a_k}}}{{{m_k}}}} \right)^{{m_i} - i}} \left[ \frac{1}{2} -
\frac{1}{\pi }\sqrt {\frac{{{a_k}b}}{{{a_k}b + {m_k}}}} \left(
\left( {\frac{\pi }{2} + {{\tan }^{ - 1}}{\omega  _1}}
\right)\sum\limits_{p = 0}^{{m_k} - i -
1} {2p \choose p} \right.\right.\notag\\
&\times\left.\left.{\left({4\left( {1 + \frac{{{a_k}b}}{{{m_k}}}}
\right)}\right)^{ - p}} + \sin \left( {{{\tan }^{ - 1}}{\omega _1}}
\right)\sum\limits_{p = 1}^{m - i - 1}{\sum\limits_{q = 1}^p
{{T_{p,q}}} } {\left( {1 + \frac{{{a_k}b}}{{{m_k}}}} \right)^{ - p}}
{\left( {\cos \left( {{{\tan }^{ - 1}}{\omega _1}} \right)} \right)^{2\left( {p - q} \right) + 1}}\right)\right]\notag\\
&-4\alpha^2 \sum\limits_{k = 0}^N {\sum\limits_{i = 0}^{{m_{_k}} -
1} {{B_{k,i}}} }\Gamma\left( {{m_k} - i} \right) {\left(
{\frac{{{a_k}}}{{{m_k}}}} \right)^{{m_i} - i}}\left[ \frac{1}{4} -
\frac{1}{\pi }\sqrt {\frac{{{a_k}b}}{{{a_k}b + {m_k}}}}
\left( \left( {\frac{\pi }{2} - {{\tan }^{ - 1}}{\omega _2}} \right)\sum\limits_{p = 0}^{{m_k} - i - 1}{2p \choose p} \right.\right.\notag\\
&\times\left.\left.{\left( {4\left( {1 + \frac{{{a_k}b}}{{{m_k}}}}
\right)} \right)^{ - p}} - \sin \left( {{{\tan }^{ - 1}}{\omega _2}}
\right)\sum\limits_{p = 1}^{m - i - 1} {\sum\limits_{q = 1}^p
{{T_{p,q}}} } {\left( {1 + \frac{{{a_k}b}}{{{m_k}}}} \right)^{ -
p}}{\left({\cos \left( {{{\tan }^{ - 1}}{\omega _2}} \right)}
\right)^{2\left( {p - q} \right) + 1}}\right)\right],
\end{align}
where ${\omega _1} = \sqrt {\frac{{{a_k}b}}{{{a_k}b + {m_k}}}} \cot
\frac{\pi }{2}$ and ${\omega _2} = \sqrt {\frac{{{a_k}b}}{{{a_k}b +
{m_k}}}} \cot \frac{\pi }{4}$.




\subsubsection{Exact Results for $G\left(\beta_{ql}\right)$}

We now analyze $G\left(\beta_{ql}\right)$ for $M$-PSK and $M$-QAM,
respectively. According to (\ref{STNC_G(beta)}), it is equivalent to
analyze ${\rm{SER}}_{ql}$. Using (\ref{STNC_source transmit 2}), the
received SNR at $R_q$ is written as
\begin{align}\label{SNR R_q}
\gamma_{ql}=\frac{{P_l \left| {h_{ql} } \right|^2 }}{{N_0 }} =
\frac{{P_l d_{ql}^{ - \alpha } }}{{N_0 }}\left| h \right|^2  =
c_{ql} \left|h\right|^2,
\end{align}
where $c_{ql} =P_l d_{ql}^{-\alpha} \slash N_0$ denotes the
equivalent SNR at $R_q$ received from $U_l$, and $h$ denotes the
unitary Nakagami-$m$ fading coefficients between $U_l$ and $R_q$
with variance one.

We first derive ${\rm{SER}}_{ql}$ for $M$-PSK as
\begin{align}\label{SER naka SER_ql}
{\rm{SER}}_{ql,{\textrm{MPSK}}}=&\frac{{\alpha(M - 1)}}{M} -
\frac{\alpha}{\pi }\sqrt {\frac{{c_{ql} b}}{{c_{ql} b +
m_{lq}}}}\left[\left( {\frac{\pi }{2} + \tan ^{ - 1} \varpi}
\right)\sum_{p = 0}^{m_{lq} - 1} {2p \choose p} \left( {4\left( {1 +
\frac{{c_{ql} b}}{m_{lq}}}
\right)} \right)^{ - p}\right.\notag\\
&+\left.\sin \left( {\tan ^{ - 1} \varpi } \right)\sum_{p =
1}^{m_{lq} - 1} {\sum_{t = 1}^p {T_{p,t} } } \left( {1 +
\frac{{c_{ql} b}}{m_{lq}}} \right)^{ - p} \left( {\cos \left( {\tan
^{ - 1} \varpi } \right)} \right)^{2\left( {p - t} \right) +
1}\right],
\end{align}
where $\varpi = \sqrt {\frac{{c_{ql} b}}{{c_{ql} b + m_{lq}}}} \cot
\frac{\pi }{M}$.

We then derive ${\rm{SER}}_{ql}$ for $M$-QAM as
\begin{align}\label{SER naka SER_ql QAM}
{\rm{SER}}_{ql,{\textrm{MQAM}}} =&2\alpha - \frac{4\alpha }{\pi
}\sqrt {\frac{{{c_{ql}}b}}{{{c_{ql}}b + {m_k}}}}\left[ \left(
{\frac{\pi }{2} + {{\tan }^{ - 1}}{\varpi _1}}
\right)\sum\limits_{p=0}^{{m_k} - i - 1} {2p \choose p}  {\left({4\left( {1 + \frac{{{c_{ql}}b}}{{{m_k}}}} \right)} \right)^{ - p}}\right.\notag\\
&+\left.\sin \left( {{{\tan }^{ - 1}}{\varpi  _1}} \right)\sum\limits_{p = 1}^{m - i - 1}{\sum\limits_{q = 1}^p {{T_{p,q}}} } {\left( {1 + \frac{{{c_{ql}}b}}{{{m_k}}}} \right)^{ - p}}
{\left({\cos \left( {{{\tan }^{ - 1}}{\varpi  _1}} \right)} \right)^{2\left( {p - q} \right) + 1}}\right]\notag\\
&-\alpha^2 + \frac{{{4\alpha ^2}}}{\pi }\sqrt
{\frac{{{a_k}b}}{{{a_k}b + {m_k}}}}\left[\left(
{\frac{\pi }{2} - {{\tan }^{ - 1}}{\varpi  _2}} \right)\sum\limits_{p = 0}^{{m_k} - i - 1}{2p \choose p} {\left({4\left( {1 + \frac{{{c_{ql}}b}}{{{m_k}}}} \right)} \right)^{ - p}}\right.\notag\\
&\left.- \sin \left( {{{\tan }^{ - 1}}{\varpi  _2}}
\right)\sum\limits_{p = 1}^{m - i - 1} {\sum\limits_{q = 1}^p
{{T_{p,q}}} } {\left( {1 + \frac{{{c_{ql}}b}}{{{m_k}}}} \right)^{ -
p}}{\left( {\cos \left( {{{\tan }^{ - 1}}{\varpi  _2}} \right)}
\right)^{2\left( {p - q} \right) + 1}}\right],
\end{align}
where ${\varpi _1} = \sqrt {\frac{{{c_{ql}}b}}{{{a_k}b + {m_{lq}}}}}
\cot \frac{\pi }{2}$ and ${\varpi _2} = \sqrt
{\frac{{{c_{ql}}b}}{{{a_k}b + {m_{lq}}}}} \cot \frac{\pi }{4}$.


Substituting (\ref{SER naka SER_ql}) and (\ref{SER naka SER_ql QAM})
into (\ref{STNC_G(beta)}), we obtain ${G\left({\beta _{ql}}\right)}$
for $M$-PSK and $M$-QAM, respectively. Therefore, we insert
(\ref{STNC_G(beta)}) and (\ref{SER naka_PSK}) into (\ref{SER naka
final}), which yields the exact closed-form SER for $M$-PSK, and
substitute (\ref{STNC_G(beta)}) and (\ref{SER naka_QAM}) into
(\ref{SER naka final}), which gives the exact SER for $M$-QAM.
Observing (\ref{SER naka_PSK}), (\ref{SER naka_QAM}), (\ref{SER naka
SER_ql}), and (\ref{SER naka SER_ql QAM}), we see that the exact SER
expressions for $M$-PSK and $M$-QAM are given in closed-form and are
valid to arbitrary numbers of sources and relays.

\subsection{Asymptotic SER}

We now provide useful insights into the network behavior in the high
SNR regime. In doing so, new compact closed-form expressions are
presented for the asymptotic SER.

We first focus on $M$-PSK. Based on (\ref{STNC_SNR S_l}),
${\rm{SER}}_{\gamma _{l|S_{l}}}$ for $M$-PSK can be alternatively
written as
\begin{align}\label{SER naka Asy 1}
&{\rm{SER}}_{\gamma_{l|S_{l}},{\textrm{MPSK}}}\notag\\
&=\frac{\alpha}{\pi}\int_0^{\left( {M - 1} \right)\pi /M}
{\int_0^\infty{\exp \left( { - \frac{{b\gamma _{l|S_{l} } }}{{\sin ^2 \theta }}} \right)f_{_{\gamma _{l|S_{l} } } } \left( {\gamma _{l|S_{l} } } \right)} d\gamma _{l|S_{l} } d\theta }\notag\\
&= \frac{\alpha }{\pi }\int_0^{\left( {M - 1} \right)\pi /M} {\left(
{\int_0^\infty  {\int_0^\infty  { \cdots \int_0^\infty  {\exp \left(
{ - \frac{{b\sum_{i = 0}^N {y_i } }}{{\sin ^2 \theta }}} \right)
\left({\prod\limits_{i = 0}^N {f_{Y_i } \left( {y_i } \right)}}
\right)\prod\limits_{i = 0}^N {dy_i } } } } } \right)d\theta }.
\end{align}
Substituting (\ref{Y_n Naka}) into (\ref{SER naka Asy 1}) and using
\cite[eq. (3.351.3)]{rvq:Table_of_Integrals}, we obtain
${\rm{SER}}_{\gamma_{l|S_{l}},{\textrm{MPSK}}}$ as
\begin{align}\label{SER naka Asy 2}
{\rm{SER}}_{\gamma_{l|S_{l}},{\textrm{MPSK}}}
&=\frac{\alpha }{\pi }\int_0^{\left( {M - 1} \right)\pi /M} {\prod\limits_{i = 0}^N {\left( {\int_0^\infty  {\frac{{{m_i}^{m_i} y_i ^{{m_i} - 1} }}{{{a_i}^{m_i} \Gamma \left( {m_i} \right)}}\exp \left( { - \left( {\frac{b}{{\sin ^2 \theta }} + \frac{{m_i}}{{a_i }}} \right)y_i } \right)dy_i } } \right)} d\theta }\notag\\
&=\frac{\alpha }{\pi }\int_0^{\left( {M - 1} \right)\pi /M}
{\prod\limits_{i = 0}^N {\left( {\frac{{{m_i}^{m_i} }}{{a_i^{m_i}
}}\left( {\frac{b}{{\sin ^2 \theta }} + \frac{{m_i}}{{a_i }}}
\right)^{ - {m_i}} } \right)} d\theta}.
\end{align}
We next use \cite[eq. (2.513.1)]{rvq:Table_of_Integrals} to develop
an asymptotic expression for
${\rm{SER}}_{\gamma_{l|S_{l}},{\textrm{MPSK}}}$ as
\begin{align}\label{SER naka Asy 3}
{\rm{SER}}_{{\gamma _{l|{S_l}}},{\rm{MPSK}}}^\infty & \le \frac{\alpha }{\pi }\int_0^{\left( {M - 1} \right)\pi /M} {\prod\limits_{i = 0}^N {\left( {\frac{{{m_i}^{{m_i}}}}{{a_i^{{m_i}}}}{{\left( {\frac{b}{{{{\sin }^2}\theta }}} \right)}^{ - {m_i}}}} \right)} d\theta } \notag\\
 &= \frac{\alpha }{\pi }\left( {\prod\limits_{i = 0}^N {{{\left( {\frac{{{m_i}}}{{{a_i}b}}} \right)}^{{m_i}}}} } \right)\int_0^{\left( {M - 1} \right)\pi /M} {{{\sin }^{2\sum\nolimits_{i = 0}^N {{m_i}} }}\theta d\theta } \notag\\
 &= \frac{\alpha }{{\pi {b^{\sum\nolimits_{i = 0}^N {{m_i}} }}}}{A_{M,N}}{\prod\limits_{i = 0}^N {\left( {\frac{{{m_i}}}{{{a_i}}}} \right)} ^{{m_i}}}
\end{align}
where
\begin{align}\label{SER naka Asy 2 note}
A_{M,N}=&\frac{1}{{2^{ 2\sum\nolimits_{i = 0}^N {{m_{i}}}} }}{{ 2\sum\nolimits_{i = 0}^N
{{m_{i}}}}\choose { \sum\nolimits_{i = 0}^N {{m_{i}}}}}\frac{\left({M - 1}\right)\pi}{
M}+\frac{{\left( { - 1} \right)^{ \sum\nolimits_{i = 0}^N {{m_{i}}}} }}{{
2^{2{ \sum\nolimits_{i = 0}^N {{m_{i}}}} - 1} }}\notag\\
&\times\sum_{k = 0}^{{ \sum\nolimits_{i = 0}^N {{m_{i}}}} - 1} {\left( { - 1} \right)^k{2{
\sum\nolimits_{i = 0}^N {{m_{i}}}}\choose k}\frac{{\sin \left( {\left( {2{ \sum\nolimits_{i = 0}^N
{{m_{i}}}} - 2k} \right)\frac{\left({M - 1}\right)\pi}{M}} \right)}}{{2{ \sum\nolimits_{i = 0}^N
{{m_{i}}}} - 2k}}}.
\end{align}
Similarly, an asymptotic expression for
${\rm{SER}}_{ql,{\textrm{MPSK}}}$ is obtained as
\begin{equation}\label{SER naka Asy SER_ql}
{\rm{SER}}_{ql,{\textrm{MPSK}}}^{\infty}\le\frac{\alpha }{\pi
}\left({\frac{{{m_{lq}} }}{{b }}}\right)^{{m_{lq}}}
\frac{A_{M,{m_{lq}},0}}{{c_{ql}^{m_{lq}}}}.
\end{equation}
Correspondingly, the asymptotic $G\left({\beta_{ql}}\right)$ for
$M$-PSK is given as
\begin{equation}\label{SER naka Asy G}
G\left({\beta_{ql}}\right)_{\textrm{MPSK}}^{\infty}=\left\{
{\begin{array}{ll}
{1,} & {{\rm{if }}~{\beta_{ql}=1}}  \\
{{\rm{SER}}_{ql,{\textrm{MPSK}}}^{\infty},} & {{\rm{otherwise.}}}  \\
\end{array}}\right.
\end{equation}
Based on (\ref{SER naka Asy 3}), (\ref{SER naka Asy SER_ql}), and
(\ref{SER naka Asy G}), the asymptotic SER for $M$-PSK is derived as
\begin{equation}\label{SER naka final up bound}
{\rm{SER}}_{l,{\textrm{MPSK}}}^{\infty}=\sum_{S_{l}=0}^{2^Q-1}{\rm{SER}}_{\gamma_{l|S_{l}},{\textrm{MPSK}}}^{\infty}
\prod_{q=1}^{Q}G\left({\beta_{ql}}\right)_{\textrm{MPSK}}^{\infty}.
\end{equation}

Following the same procedure outlined for $M$-PSK, we derive the
asymptotic SER for $M$-QAM as
\begin{equation}\label{SER naka final up bound_QAM}
{\rm{SER}}_{l,{\textrm{MQAM}}}^{\infty}=\sum_{S_{l}=0}^{2^Q-1}{\rm{SER}}_{\gamma_{l|S_{l}},{\textrm{MQAM}}}^{\infty}
\prod_{q=1}^{Q}G\left({\beta_{ql}}\right)_{\textrm{MQAM}}^{\infty},
\end{equation}
where the asymptotic ${\rm{SER}}_{\gamma_{l|S_{l}},}$ for $M$-QAM is
derived as
\begin{align}\label{SER naka Asy QAM}
{\rm{SER}}_{\gamma_{l|S_{l}},{\textrm{MQAM}}}^{\infty}\le \frac{4 \alpha }{{\pi {b^{\sum\nolimits_{i =
0}^N {{m_i}} }}}}{A_{2,N}}{\prod\limits_{i = 0}^N {\left( {\frac{{{m_i}}}{{{a_i}}}} \right)} ^{{m_i}}}
- \frac{{4{\alpha ^2}}}{{\pi {b^{\sum\nolimits_{i = 0}^N {{m_i}} }}}}{A_{4/3,N}}{\prod\limits_{i =
0}^N {\left( {\frac{{{m_i}}}{{{a_i}}}} \right)} ^{{m_i}}},
\end{align}
with  ${A_{2,{m_i},N}}$ and ${A_{4/3,{m_i},N}}$ being defined in (\ref{SER naka Asy 2 note}), and the
asymptotic $G\left({\beta_{ql}}\right)$ for $M$-QAM is derived as
\begin{equation}\label{SER naka Asy G_QAM}
G\left({\beta_{ql}}\right)_{\textrm{MQAM}}^{\infty}=\left\{
{\begin{array}{ll}
{1,} & {{\rm{if }}~{\beta_{ql}=1}}  \\
{{\rm{SER}}_{ql,{\textrm{MQAM}}}^{\infty},} & {{\rm{otherwise,}}}\\
\end{array}} \right.
\end{equation}
with
\begin{equation}\label{SER naka Asy SER_ql QAM}
{\rm{SER}}_{ql,{\textrm{MQAM}}}^{\infty}\le\frac{{4\alpha }}{\pi
}{\left( {\frac{{{m_{lq}}}}{b}}
\right)^{{m_{lq}}}}\frac{{{A_{2,{m_{lq}},0}}}}{{c_{ql}^{{m_{lq}}}}}-
\frac{{4{\alpha ^2}}}{\pi }{\left( {\frac{{{m_{lq}}}}{b}}
\right)^{{m_{lq}}}}\frac{{{A_{4/3,{m_{lq}},0}}}}{{c_{ql}^{{m_{lq}}}}}.
\end{equation}

Based on  (\ref{SER naka final up bound}) and (\ref{SER naka final up bound_QAM}), we next examine the
diversity order of the network, which represents the slope of the SER against average SNR in a log-log
scale. According to (\ref{SER naka final up bound}) and (\ref{SER naka final up bound_QAM}), the
asymptotic SER of $U_l$ can rewritten as
\begin{equation}\label{SER HIGH naka}
{\rm{SER}}_l^\infty  = \sum\limits_{{S_l} = 0}^{{2^Q} - 1} {{\Theta
_{{S_l}}}\left[ {\prod\limits_{i = 0}^N {\frac{1}{{a_i^{{m_i}}}}} }
\right]} \prod\limits_{q = 1}^Q {\frac{1}{{c_{ql}^{{m_{lq}}}}}},
\end{equation}where $\Theta_{{S_l}}$ denotes the coefficient independent of $a_i$ and $c_{ql}$.in which the diversity order can be confirmed as
\begin{equation}\label{diversity order naka}
{\rm{div}}={m_{ld}}+\sum\nolimits_{q=1}^Q{\min\left({{m_{qd}},{m_{lq}}}\right)}.
\end{equation}
It is evident from (\ref{diversity order naka}) that the full
diversity order is achieved, which is determined by the Nakagami-$m$
fading parameters of all the channels. Notably, the diversity order
is independent of the number of sources. In particular, this full
diversity order is preserved even non-orthogonal STNC codes are
employed.




\section{Simulation and Numerical Results}

In this section, simulation and numerical results are presented to
examine the impact of network parameters with STNC (e.g., the number
of relays, the relay location, Nakagami-$m$ fading parameters, and
power allocation) on the SER of $U_l$. In the figures, we consider a
practical scenario where the relays are placed at different
distances from $D$ and $U_l$ with $c_j \ne c_i$ and $c_{jl} \ne
c_{il}$ for $j \ne i$. We set the distance between $U_l$ and $D$ as
$d_{ld}=1$. The cross correlations between different spread codes,
defined in (\ref{STNC_source transmit 1}), are set to be zero. We
also assume equal transmit power at each node. Further, our results
concentrate on the practical example of a highly shadowed area with
the path loss exponent as $\alpha=3.5$ \cite{rvq:david}. In the
figures, the exact SER for $M$-PSK is evaluated by substituting
(\ref{STNC_G(beta)}), (\ref{SER naka_PSK}), and (\ref{SER naka
SER_ql}) into (\ref{SER naka final}), and the exact SER for $M$-QAM
is evaluated by substituting (\ref{STNC_G(beta)}), (\ref{SER
naka_QAM}), and (\ref{SER naka SER_ql QAM}) into (\ref{SER naka
final}). The asymptotic SER for $M$-PSK and $M$-QAM is calculated
from (\ref{SER naka final up bound}) and (\ref{SER naka final up
bound_QAM}), respectively.

\subsection{Impact of Number of Relays and Equal Nakagami-$m$ Fading Parameters}

\begin{figure}[!t]
\centering
\includegraphics[width=4in]{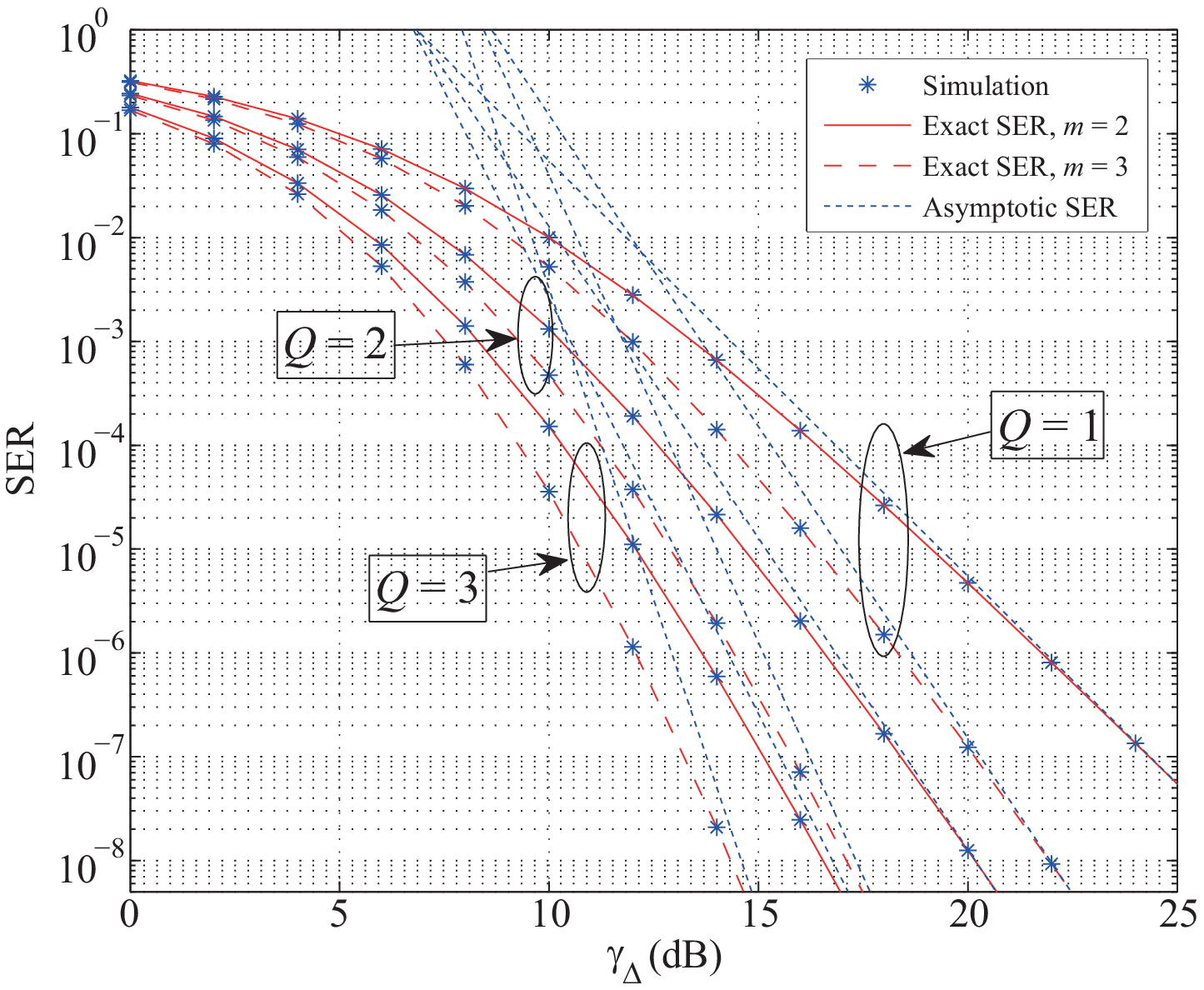}
\caption{Exact and asymptotic SER with 4QAM for $Q=1, 2, 3$,
$c_{\left({i+1}\right)l}=c_{i+1}=c_{i}+0.1 \gamma_{\Delta}$, and
$c_0=c_{0l}=0.6 \gamma_{\Delta}$.} \label{QPSK.eps}
\end{figure}

\begin{figure}[!t]
\centering
\includegraphics[width=4in]{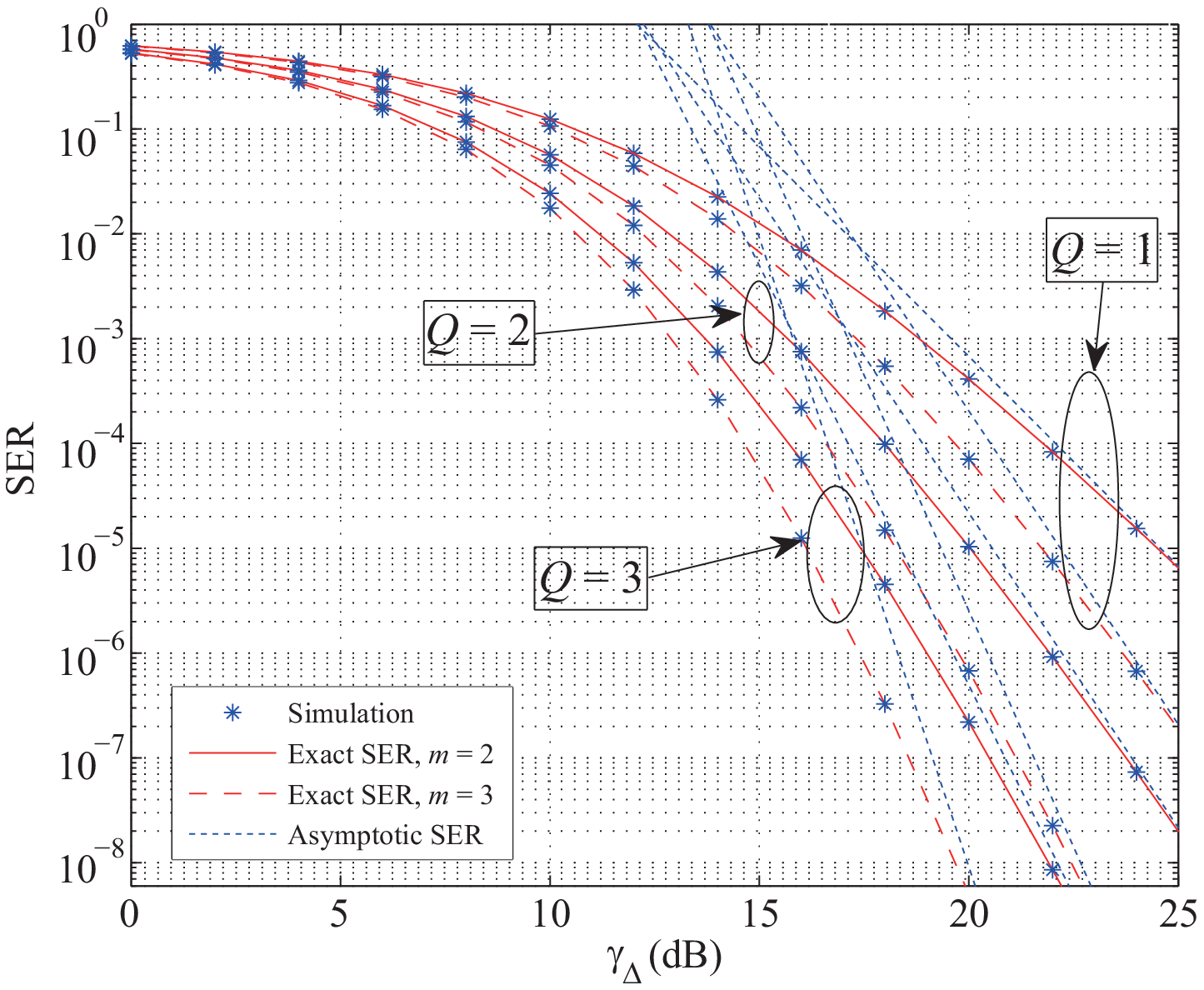}
\caption{Exact and asymptotic SER with 8PSK for $Q=1, 2, 3$,
$c_{\left({i+1}\right)l}=c_{i+1}=c_{i}+0.1 \gamma_{\Delta}$, and
$c_0=c_{0l}=0.6 \gamma_{\Delta}$.} \label{8PSK.eps}
\end{figure}

In this subsection, we focus on equal Nakagami-$m$ fading parameters
with $m_{i}=m$. The average received SNRs are set as
$c_{\left({i+1}\right)l}=c_{i+1}=c_{i}+0.1\gamma_{\Delta}$ and
$c_0=c_{0l}=0.6\gamma_{\Delta}$. Fig. \ref{QPSK.eps} plots the exact
and asymptotic SER with 4QAM. Fig. \ref{8PSK.eps} plots the exact
and asymptotic SER with 8PSK. From Figs. \ref{QPSK.eps} and
\ref{8PSK.eps}, we see that the asymptotic SER curves accurately
predict the exact ones in the high SNR regime. By observing these
asymptotic curves, it is evident that the diversity order increases
with $Q$, which indicates that increasing the number of relays
brings an improved performance. It is also seen that the diversity
order increases with $m$, which indicates that the improvement in
fading channels leads to a reduction in the SER. Moreover, we see
that the simulation points are in precise agreement with our exact
analytical curves, which demonstrates the correctness of our
analysis in Section III. Comparing the SER in Fig. \ref{QPSK.eps}
with that in Fig. \ref{8PSK.eps}, we further see a poorer network
performance is achieved by higher order modulation schemes.




\subsection{Impact of Relay Location}

\begin{figure}[!t]
\centering
\includegraphics[width=4in]{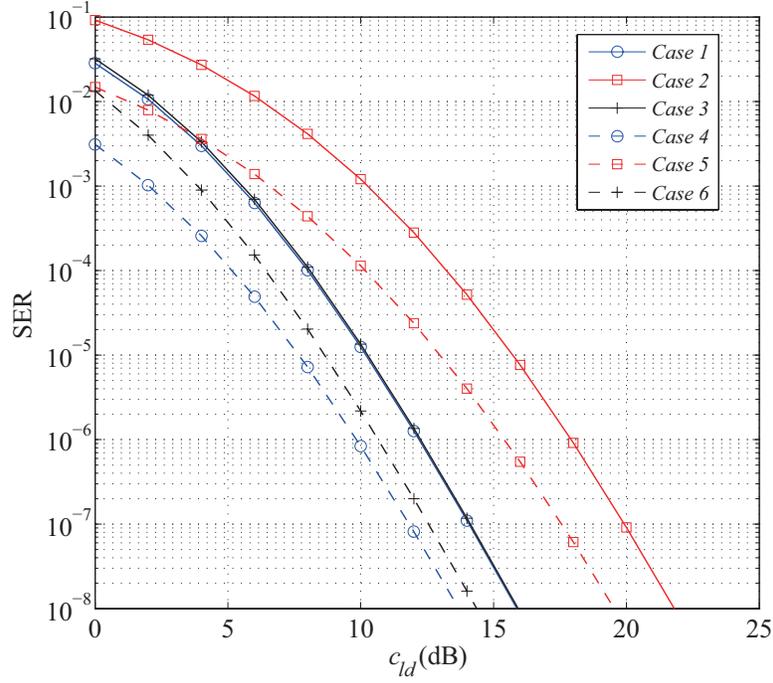}
\caption{Exact SER with BPSK for $Q=2$, $m=2$ and 6 cases:
\emph{Case 1}: $d_{l1}=0.2,d_{l2}=0.3,d_{1d}=1.1,d_{2d}=1.2$;
\emph{Case 2}: $d_{l1}=0.2,d_{l2}=0.3,d_{1d}=2,d_{2d}=2.2$;
\emph{Case 3}: $d_{l1}=0.8,d_{l2}=0.9,d_{1d}=1.1,d_{2d}=1.2$;
\emph{Case 4}: $d_{l1}=1.1,d_{l2}=1.2,d_{1d}=0.2,d_{2d}=0.3$;
\emph{Case 5}: $d_{l1}=2,d_{l2}=2.2,d_{1d}=0.2,d_{2d}=0.3$; and
\emph{Case 6}:
$d_{l1}=1.1,d_{l2}=1.2,d_{1d}=0.8,d_{2d}=0.9$.}\label{relay_location.eps}
\end{figure}

In this subsection, we consider ${d_{lq}} \ne {d_{qd}}$, which leads
to ${c_{q}} \ne {c_{qd}}$, and consider equal Nakagami-$m$ fading
parameters with $m_{i}=m=2$. We further normalize ${d_{ld}}$ to
unity with ${d_{ld}}=1$. Fig. \ref{relay_location.eps} plots the
exact SER with BPSK for $Q=2$. In this figure, \emph{Cases 1},
\emph{2}, \emph{3} represent the scenario where the relays are
located close to the source, while \emph{Cases 4}, \emph{5},
\emph{6} represent the scenario where the relays are located close
to the destination.


We first consider \emph{Cases 1}, \emph{2}, and \emph{3}. We see
that \emph{Case 1} offers a prominent SNR advantage relative to
\emph{Case 2}. This indicates that the reduction in the distance
between the relay and the destination brings a substantial SER
improvement. We also see that \emph{Case 1} and \emph{Case 3}
achieve almost the same SER across the entire SNR range. This
indicates that the SER improvement from the reduced distance between
the source and the relay is negligible. These observations are due
to the fact that the network performance is dominant by the
relay-destination link when the relays are close to the source. As
such, the quality improvement of the relay-destination link has a
higher positive impact on the SER than that of the source-relay
link.

We next consider \emph{Cases 4}, \emph{5}, and \emph{6}. It is seen
that \emph{Case 4} provides a substantial SNR advantage compared to
\emph{Case 5}. It is also seen that \emph{Case 4} achieves a slight
SNR advantage compared to \emph{Case 6}. These observations are
explained by the fact that the network performance is dominant by
the source-relay link when the relays are close to the destination.

\subsection{Impact of Unequal Nakagami-$m$ Fading Parameters}

\begin{figure}[!t]
\centering
\includegraphics[width=4in]{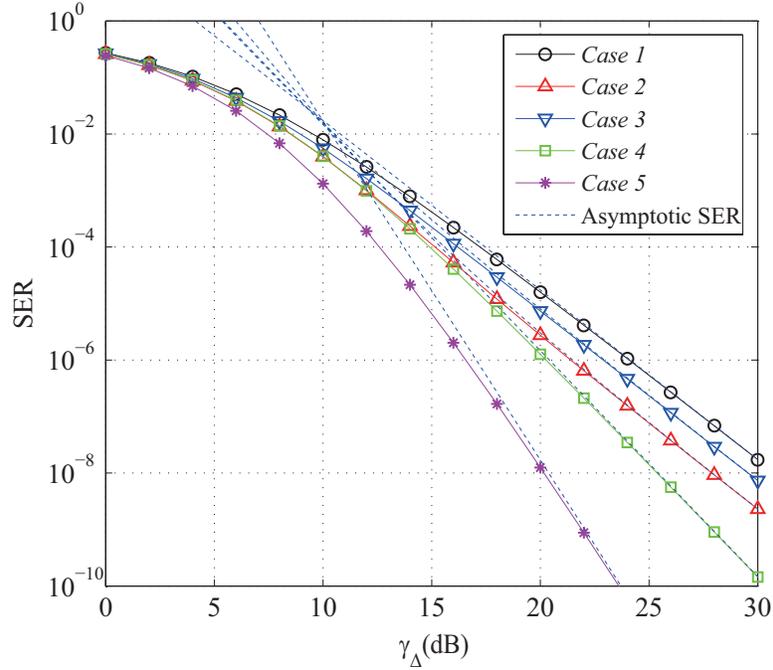}
\caption{Exact SER with 4QAM for $Q=2$,
$c_{\left({i+1}\right)l}=c_{i+1}=c_{i}+0.1 \gamma_{\Delta}$,
$c_0=c_{0l}=0.6 \gamma_{\Delta}$ and 6 cases: \emph{Case 1}:
$m_i=m=1$; \emph{Case 2}: $m_{l1}=1, m_{l2}=1, m_{ld}=1, m_{1d}=2,
m_{2d}=2$; \emph{Case 3}: $m_{l1}=2, m_{l2}=2, m_{ld}=1, m_{1d}=1,
m_{2d}=1$; \emph{Case 4}: $m_{l1}=1, m_{l2}=2, m_{ld}=1, m_{1d}=1,
m_{2d}=2$; \emph{Case 5}: $m_i=m=2$.} \label{QPSK_DUO_GE_M.eps}
\end{figure}

We concentrate on unequal Nakagami-$m$ fading parameters and set the
average received SNRs as
$c_{\left({i+1}\right)l}=c_{i+1}=c_{i}+0.1\gamma_{\Delta}$ and
$c_0=c_{0l}=0.6\gamma_{\Delta}$. Fig. \ref{QPSK_DUO_GE_M.eps} plots
the exact SER with 4QAM for $Q=2$. This figure clearly shows that
the diversity order in (\ref{diversity order naka}) is accurate. For
example, it is evident that the asymptotic SER curves of \emph{Cases
1}, \emph{2}, and \emph{3} are in parallel, which indicates that
they achieve the same diversity order. As indicated in
(\ref{diversity order naka}), \emph{Cases 1}, \emph{2}, and \emph{3}
achieve identical diversity order of $3$. Moreover, we see that the
diversity order of \emph{Case 4} increases to $4$, and the diversity
order of \emph{Case 5} increases to $6$. This is predicted by
(\ref{diversity order naka}), which shows that the diversity order
is determined by the Nakagami-$m$ fading parameters of all the
channels.

\subsection{Impact of Power Allocation}

\begin{figure}[!t]
\centering
\includegraphics[width=4in]{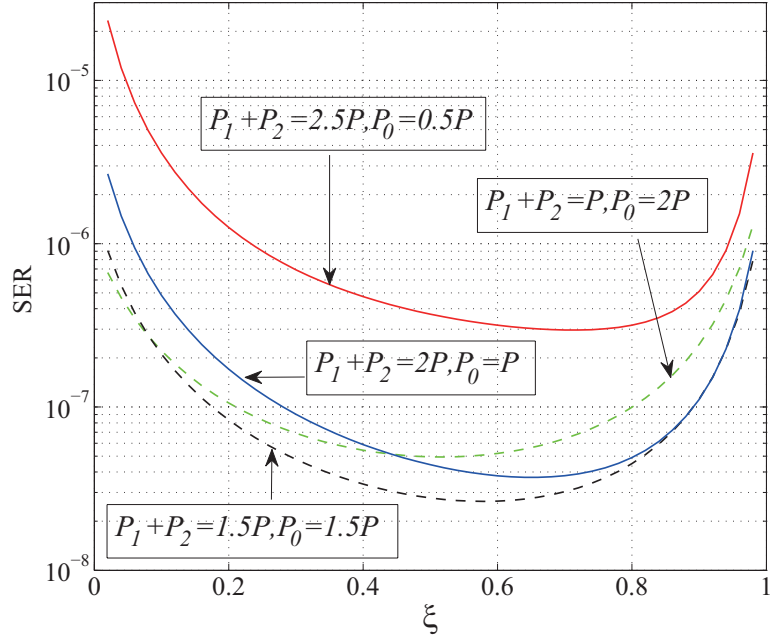}
\caption{Exact SER with 4QAM for $Q=2$, $m=2$ with ${d_{l1}} = 0.8$, ${d_{l2}} = 1$, ${d_{1d}} = 0.9$,
${d_{2d}}=0.7$, and different power allocation $P_1$, $P_2$, $P_3$.} \label{power_location.eps}
\end{figure}

We now focus on arbitrary transmit power at each node. We consider
equal Nakagami-$m$ fading parameters with $m_{i}=m=2$, set the relay
location as ${d_{l1}} = 0.8$, ${d_{l2}} = 1$, ${d_{1d}}=0.9$, and
${d_{2d}}=0.7$, and normalize ${d_{ld}}$ as ${d_{ld}}=1$. We denote
the transmit powers at $U_l$, $R_1$, and $R_2$ as $P_0$, $P_1$, and
$P_2$, respectively. Under the total power constraint, we have
$P_0+P_1+P_2=3P$. Fig. \ref{power_location.eps} plots the exact SER
versus $\xi=P_1/(P_1+P_2)$ with 4QAM for $Q=2$ and
$P/N_0=12~{\rm{dB}}$. We see that the optimal value of $P_1/P_2$
depends on $P_0$. For example, the optimal power allocation is at
$P_1=1.75P$ and $P_2= 0.75P$ when $P_0=0.5P$. Moreover, the optimal
power allocation is at $P_1=P_2= 0.5P$ when $P_0=2P$. Using our SER
expressions with different relay locations, iterative search method
can be used to find the optimal power allocation that minimizes the
SER.




\subsection{Impact of Nonorthogonal Codes}

\begin{figure}[!t]
\centering
\includegraphics[width=4in]{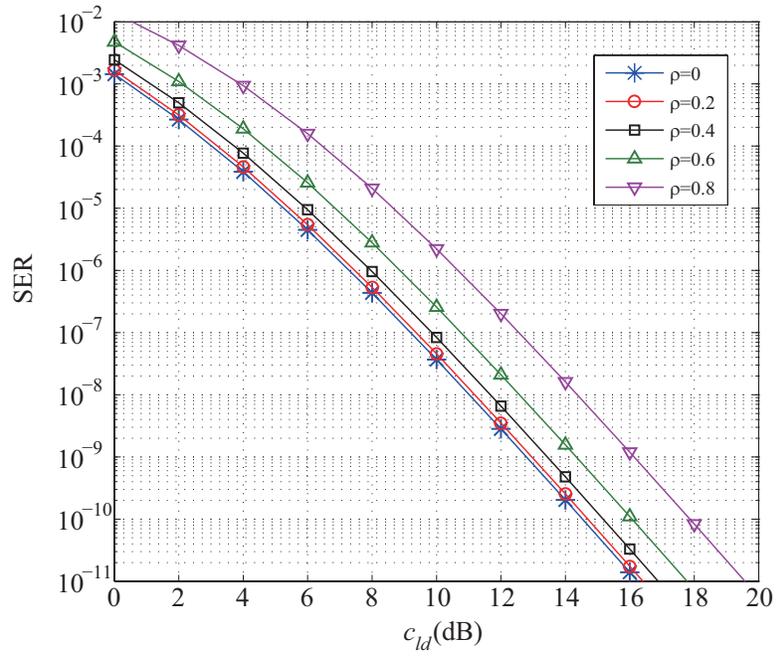}
\caption{Exact SER with 4QAM for $Q=2$, $N=3$, $m_{i}=2$ and different cross correlation $\rho$ with
${d_{l1}} = 0.6$, ${d_{l2}} = 0.8$, ${d_{1d}} = 0.6$ and ${d_{2d}}=0.8$, where ${d_{ld}}$ is
normalized as ${d_{ld}}=1$.} \label{cross_factor.eps}
\end{figure}

We now turn our attention to the impact of nonorthogonal codes. We
consider $\rho_{pq}=\rho\ne0$ for all $p,q$ and equal Nakagami-$m$
fading parameters with $m_{i}=2$. Fig. \ref{cross_factor.eps} plots
the exact SER with 4QAM for $Q=2$ and $N=3$. The case of $\rho=0$
represents orthogonal codes. We see a reduction in the SER as $\rho$
increases. We also see that the diversity order is not affected by
cross correlation. As such, the nonorthogonal codes which permit
broader applications can be used for higher throughput without
sacrificing the error rate significantly.


\section{Conclusions}

In this paper, we analyzed the SER of STNC in a distributed
cooperative network where $L$ sources communicate with a single
destination with the assistance of $Q$ relays. For $M$-PSK and
$M$-QAM modulation, new exact closed-form expressions of SER over
independent but not necessarily identically distributed Nakagami-$m$
fading channels were derived. Moreover, the asymptotic SER was
derived to reveal the network performance in the high SNR regime.
Specifically, the asymptotic SER reveals that the diversity order of
STNC was determined by the Nakagami-$m$ fading parameters of all the
channels. Simulation results were used to validate our analytical
expressions and to examine the impact of Nakagami-$m$ fading
parameters, relay location, power allocation, and nonorthogonal
codes on the SER.

\appendices

\section{Proof of (\ref{residue of all pole naka})}\label{app_Residue}

\begin{figure}[!t]
\centering
\includegraphics[width=4in]{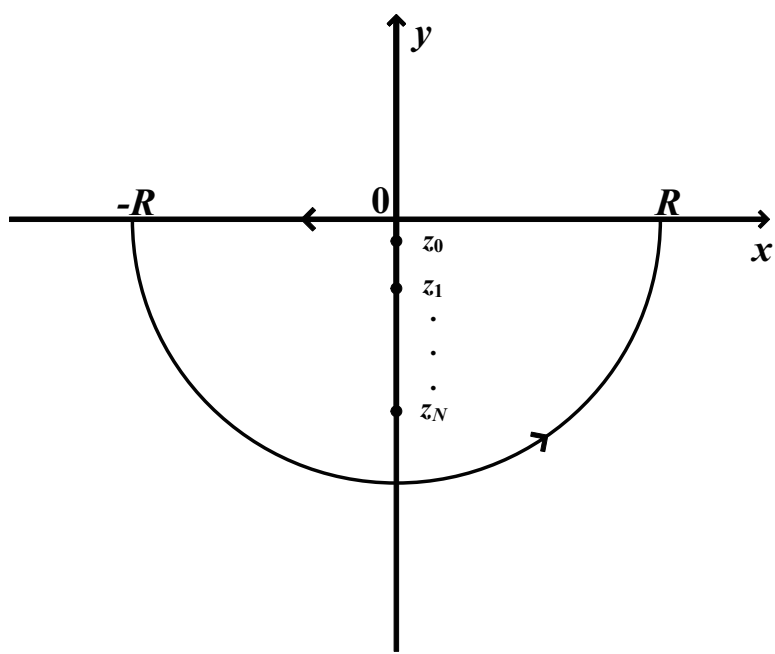}
\caption{The distribution of the poles.} \label{residue.eps}
\end{figure}

According to Fig. \ref{residue.eps}, all poles are located in a closed curve. According to the residue
theorem, the integral over a closed curve can be expressed as the linear combination of the residues
of the poles in the curve. Mathematically, we have
\begin{equation}\label{residue theorem 1 naka}
\int_R^{ - R} {g\left( u \right)} du + \int_{C_R}{g\left(z\right)dz} = 2\pi j\sum_{k = 0}^N {\rm
Res}\left[ {g\left( {z_k } \right),z_k } \right],
\end{equation}where $C_R$ is the counterclockwise semicircular curve
in Fig. 3. The absolute value of $\int_{C_R}{g\left(z\right)dz}$ in (\ref{residue theorem 1 naka}) can
be upper bounded as
\begin{align}\label{residue theorem 2 naka}
\left| {\int_{C_R } {g\left( z \right)dz} } \right| &\le \int_{C_R } {\left| {g\left( z \right)} \right|dz}\notag\\
& = \int_{C_R } {\left| {\left( {\prod\limits_{n = 0}^N {\left( {1 - \frac{{jza_n }}{m_n}} \right)^{ - m_n} } } \right)e^{ - jzv} } \right|dz}\notag\\
&\le \int_{C_R } {\left| {\prod\limits_{n = 0}^N {\frac{1}{{\left( {1 - ja_n z} \right)^{m_n} }}} } \right|\left| {e^{ - jzv} } \right|dz}\notag\\
&= \int_{C_R } {\prod\limits_{n = 0}^N {\left| {\frac{1}{{\left( {1 - ja_n z} \right)^{m_n} }}}
\right|} dz}.
\end{align}
When $R \to \infty$, which indicates that the integral range of $g\left({u}\right)$ is from $\infty $
to $-\infty$, we find the property of (\ref{residue theorem 2 naka}) as
\begin{align}\label{abs 1 naka}
\left| {\int_{C_R } {g\left( z \right)dz} } \right| &= \int_{C_R } {\prod\limits_{n = 0}^N {\frac{1}{{{a_n}^{m_n} \left| z \right|^{m_n}  + O\left( {z^{m_n}  } \right)}}} dz}\notag\\
&= \left( {\prod\limits_{n = 0}^N {\frac{1}{{{a_n}^{m_n} }}} } \right)\int_{C_R } {\frac{1}{{\left| z \right|^{\sum\nolimits_{n = 0}^N {{m_n}} } }}dz}\notag\\
&= \left( {\prod\limits_{n = 0}^N {\frac{1}{{{a_n}^{m_n} }}} } \right)\frac{\pi }{{R^{\sum\nolimits_{n = 0}^N {{m_n}}  - 1} }}\notag\\
&\to 0.
\end{align}
Substituting (\ref{abs 1 naka}) into (\ref{residue theorem 1 naka}), we obtain (\ref{residue of all
pole naka}) and thus complete this proof.

\bibliographystyle{IEEEtran}
\bibliography{myref}

\newpage

\end{document}